\documentclass[aps,prb,twocolumn,superscriptaddress,amsmath,amssymb]{revtex4-2}
\usepackage{newtxtext,newtxmath}
\usepackage{graphicx}
\usepackage{bm}
\usepackage{siunitx}
\usepackage{booktabs}
\usepackage{tabularx}
\usepackage[colorlinks=true,citecolor=blue,breaklinks=true,linkcolor=blue,
linktocpage=true,urlcolor=blue,pagebackref=false]{hyperref}
\usepackage{appendix}
\usepackage{titlesec}

\newcommand{\tr}{\ensuremath{\operatorname{tr}}}

\renewcommand{\paragraph}[1]{{\ignorespaces\bigskip\noindent\textbf{\textsf{#1}}\newline}}
\renewcommand{\subparagraph}[1]{{\medskip\noindent\textbf{#1.}\ }}

\newcommand{\void}[1]{}

\newcommand{\SIrefs}{}
\newcommand{\supplement}{Supplementary Material \cite{supplement}\nocite{\SIrefs}}

\begin{document}
\title{Temperature-dependent broadening of coherent current peaks in InAs double quantum dots}

\author{Olfa Dani}
\affiliation{Institut f\"ur Festk\"orperphysik, Leibniz Universit\"at Hannover, D-30167 Hanover, Germany}

\author{Robert Hussein}
\affiliation{Institut f\"ur Festk\"orpertheorie und -optik, Friedrich-Schiller-Universit\"at Jena, D-07743 Jena, Germany}

\author{Johannes C. Bayer}
\affiliation{Institut f\"ur Festk\"orperphysik, Leibniz Universit\"at Hannover, D-30167 Hanover, Germany}

\author{Sigmund Kohler}
\affiliation{Instituto de Ciencia de Materiales de Madrid, CSIC, E-28049 Madrid, Spain}

\author{Rolf J. Haug}
\affiliation{Institut f\"ur Festk\"orperphysik, Leibniz Universit\"at Hannover, D-30167 Hanover, Germany}

\date{\today}

\begin{abstract}
Quantum systems as used for quantum computation or quantum sensing are
nowadays often realized in solid state devices as e.g.\ complex Josephson
circuits or coupled quantum-dot systems. Condensed matter as an environment
influences heavily the quantum coherence of such systems. Here,
we investigate electron transport through asymmetrically coupled  
InAs double quantum dots and observe an extremely strong temperature
dependence of the coherent current peaks of single-electron tunneling. We analyze experimentally and theoretically the
broadening of such coherent current peaks up to temperatures of \SI{20}{K} and we are able to model it with quantum
dissipation being due to two different bosonic baths. These bosonic baths
mainly originate from substrate phonons. Application of a magnetic
field helps us to identify the different quantum dot states through their
temperature dependence.
\end{abstract}

\maketitle

\paragraph{Introduction}
The building blocks of quantum information technology and future quantum
computers are qubits. Qubits can be based on coherent superpositions in double
quantum dots (DQD) and such DQDs can be easily formed in a
variety of semiconducting materials. The use of semiconductor technology
guarantees more or less the necessary scalability of qubit structures.
Along these lines it has been shown recently that qubits based on quantum dots
can be formed and manipulated in CMOS technology \cite{XueNature2021a} at
not very low temperatures \cite{YangNature2020a}, i.e.\ quantum-dot
based quantum computers seem to be within reach. The first observation of a
coherent mode in a DQD system dates back more than 20 years
\cite{BlickPRL1998a}, while successful coherent manipulation of electronic states
\cite{HayashiPRL2003a,PettaPRL2004a} or of spin states
\cite{KoppensFHScience2005a,PettaScience2005a} in DQDs has been shown few years
later opening the path towards quantum information processing with quantum
dots.

Coherence properties of a quantum state depend on the influence of the
environment. Already in the early studies of DQDs it became clear that they
interact with the environment via the emission of phonons
\cite{FujisawaScience1998a,BrandesPRL1999a,WielRMP2002a}. Measurements of phonon emission were repeated in more detail just recently
\cite{HofmannPRResearch2020a}. Whereas in these works coupling to the phonon bath has
been studied in great detail for detuned quantum dots, corresponding
studies just at the resonance are scarce. In addition to the mentioning of
some temperature dependence of the so-called elastic peak in
refs.~\cite{FujisawaScience1998a,WielRMP2002a}, a theoretical work studied
phonon decoherence in 2005 \cite{VorojtsovPRB2005a}, and it has been shown that
at low temperatures electron-phonon scattering can enhance the current
noise close to resonance, as has been discussed experimentally
\cite{BartholdPRL2006a} and theoretically \cite{KiesslichPRL2007a,
BraggioPhysicaE2008a, SanchezPRB2008a}. 

Here, we focus on the detailed temperature dependence of the resonant
tunnel current which is mainly caused by the coupling to the phonon
environment. We describe the quantum dissipation of the coherent current
peaks up to temperatures of 20 K by introducing couplings to two different bosonic baths.

\begin{figure*}[t]
\centerline{\includegraphics[]{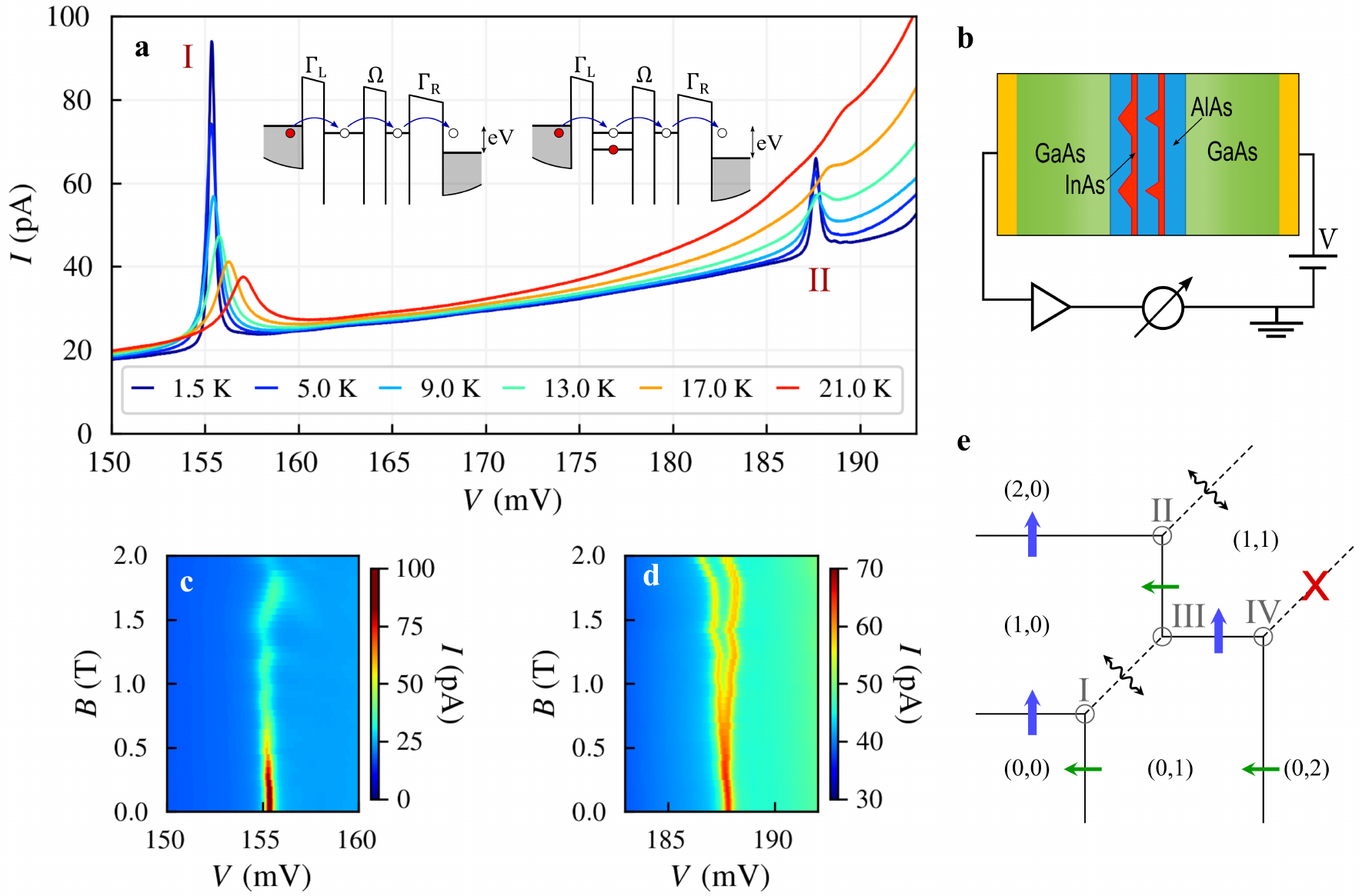}}
\caption{Experiment.
\textbf{a}, Current-voltage characteristic of InAs
double quantum dots (DQDs) for different temperatures between 1.5 and \SI{21}{K}.
The inset shows simplified energy level diagrams of the DQD system: the
left diagram for the first resonance peak (peak I) around \SI{155}{mV} and the
right diagram for the second  resonance peak (peak II) around \SI{188}{mV}.
\textbf{b}, Schematic image of the investigated heterostructure and the
measurement set-up, see methods section for detailed information.
\textbf{c}, Current of resonance I as function of voltage and magnetic
field at $T = \SI{1.5}{K}$.
\textbf{d}, Same for resonance II.
\textbf{e}, Charging diagram of the double quantum dot system with all the
triple points involving one and two electron states. The different
couplings to the two leads are depicted by arrows with different colours and
thicknesses.
}\label{fig.:scheme}
\end{figure*}

\paragraph{Experiment}
For our studies we use self-assembled InAs quantum dots similar to the ones used in Refs. \cite{BartholdPRL2006a,KiesslichPRL2007a} where the second
quantum dot grows on top of the first dot due to strain fields induced by
the InAs islands \cite{XiePRL1995a}. The second dot is slightly larger than
the first one \cite{EiseleAPL1999a}.  AlAs layers of nominally identical
thickness are used to separate the InAs quantum dots from the doped GaAs
layers and from each other, as depicted in Fig.~\ref{fig.:scheme}b. Due to
the difference in size of the two quantum dots the effective thickness of
the AlAs barriers is different and the coupling to the leads is asymmetric.
Each characteristic feature in the measured current of our device is expected to arise
from a single DQD channel although several dots are present.
The line plot in Fig.~\ref{fig.:scheme}a shows the current $I$ through the
DQD device as a function of the bias voltage $V$ for different temperatures ranging
from \SI{1.5}{\kelvin} to \SI{21}{\kelvin}. The graph shows two distinct
current peaks at $V \approx \SI{155}{\milli\volt}$ (peak I) and $V \approx
\SI{187}{\milli\volt}$ (peak II) which are due to resonant tunneling of single 
electrons through the InAs double quantum dots. The left peak (see also
left schematic level diagram in Fig.~\ref{fig.:scheme}a) corresponds to only
a single electron being present in the DQD and originates from tunnel cycles with the
occupation $(0,0) \to (1,0) \to (0,1)$. This situation is depicted in the charging diagram in
Fig.~\ref{fig.:scheme}e as triple point I. The right peak in Fig.~\ref{fig.:scheme}a
corresponds to single-electron tunneling through InAs quantum dots with an 
additional electron being present and including double occupation, namely
$(1,0) \to (2,0) \to (1,1)$ (see right schematic level diagram in
Fig.~\ref{fig.:scheme}a and triple point II in the charging diagram in
Fig.~\ref{fig.:scheme}e). Even though in a simplified understanding of
single-electron tunneling through DQDs, the tunnel resonances should not be affected by the 
Fermi distributions in the leads due to the low temperatures and the
resonances being far away from the Fermi levels in the leads, we observe
quite a strong temperature dependence of both the amplitude and the width
of both peaks in Fig.~\ref{fig.:scheme}a. With increasing temperature, the
current resonances significantly broaden and at the same time the amplitude
of the peaks decreases. Both effects are accompanied by a shift of the peak
position toward slightly more positive voltages. This shift
is attributed to temperature-dependent changes in the electric field
distribution in the sample.

Whilst both resonances show similar behaviour as function of temperature,
the magnetic field dependence reveals major differences. The colour graphs
in Fig.~\ref{fig.:scheme}c,d show the current of the respective resonance
as function of bias voltage and magnetic field up to $B=\SI{2}{\tesla}$ at
$T=\SI{1.5}{\kelvin}$.  For peak I, the main effect of the magnetic field
seems to be a reduction of the peak amplitude. The magnetic field was
applied perpendicular to the layer structure and parallel to the current.
Therefore the weak oscillation observed for the magnetic field dependence
of the resonances originates from the Landau-level structure in the
emitter. For peak II (Fig.~\ref{fig.:scheme}d), not only is the amplitude
much less affected in comparison to the first peak, but the single
resonance splits into two. This indicates that the two peaks emerge from
different electron configurations, where the left peak I corresponds to a
single-electron triple point, whereas the right peak II involves a double
occupation of the larger quantum dot. In the charging diagram in
Fig.~\ref{fig.:scheme}e two more triple points appear which are not
expected to be observable in our structure. Triple point IV will be spin
blocked, whereas triple point III corresponds to hole like transport which
exchanges the role of the asymmetric couplings and therefore, these current
peaks will be much smaller.

\begin{figure}[t]
\centerline{\includegraphics[]{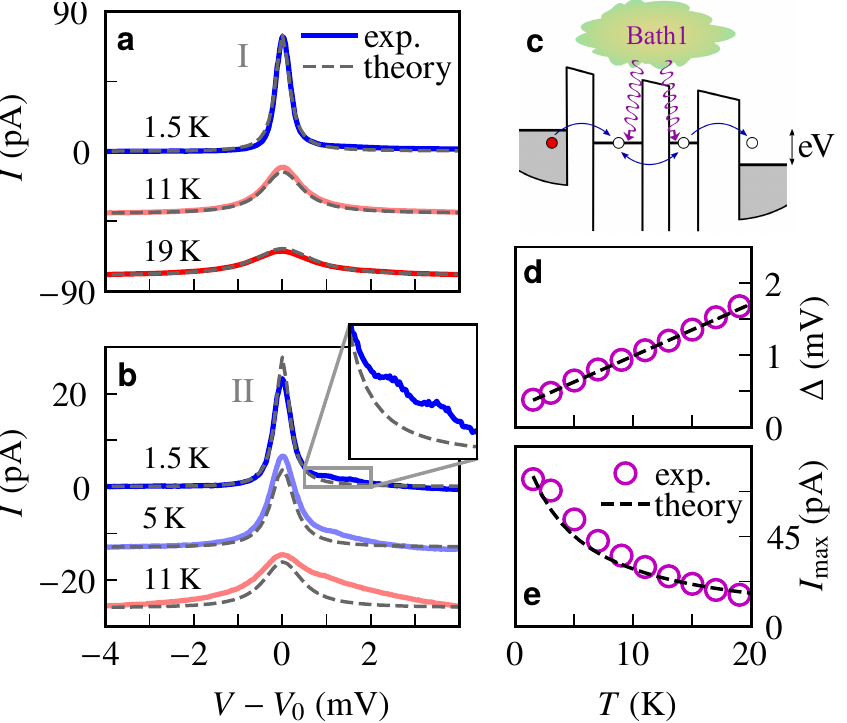}}
\caption{Analysis of the current peaks.
\textbf{a,b}, Enlargement of the measured peaks at
$V_0\approx\SI{155}{\milli\volt}$ (peak I) and $V_0\approx\SI{188}{\milli\volt}$ (peak II),
respectively, with the background subtracted (solid lines) in comparison
with the result of the theoretical prediction with only bath~1 (dashed) for
various temperatures.  The inset of panel b reveals the presence of two
further resonances.
\textbf{c}, Corresponding model with an environment coupled to the
difference of the onsite energies.
\textbf{d,e}, 
Width $\Delta$ and height $I_{\rm max}$ of peak I in dependence of the temperature.
Experimental values are depicted by circles while the dotted line shows the
current of the full model with $\Gamma_L=\SI{500}{\micro eV}$,
$\Gamma_R=\SI{30}{\micro eV}$, $\Omega=\SI{4}{\micro eV}$, $\alpha =
0.011$, and $\alpha'=0$.}
\label{fig:broadening}
\end{figure}

For now, we want to focus on the temperature dependence of the resonances.
However, not only the peaks in the current are influenced by increasing
temperature, also the off-resonant background current experiences a
temperature dependent increase.  Especially for the right peak II and the
highest temperatures the increasing background current interferes with the
appearance and the visibility of the resonance peak.  In order
to analyse the current resonances in more detail, we subtracted this
temperature-dependent background and will discuss it later.  In
Fig.~\ref{fig:broadening} the current resonances are presented after
subtraction of the background current and after normalizing to the peak position. 
For three
temperatures, peak I is presented in Fig.~\ref{fig:broadening}a, whereas in Fig.~\ref{fig:broadening}b peak II is
shown. For both peaks one sees a clear decrease in amplitude and increase
in broadening with temperature.

\paragraph{Theoretical model}
For a theoretical description, we model each quantum dot as a single
orbital with energy $\epsilon_\ell$ ($\ell=L,R$) and tunnel coupling
$\Omega$.  For states with more than one electron, we consider the onsite
and nearest-neighbour interaction energies $U$ and $U'$, respectively.
Each dot is tunnel coupled also to a lead which enables electron
transitions with the rates $\Gamma_\ell$ from lead $\ell=L,R$ to the dot
and back, depending on the Fermi energy of the respective lead.  For
sufficiently small tunnel coupling, the leads are eliminated within a
Bloch-Redfield approach \cite{RedfieldIBMJRD1957a, Breuer2003a} which leads
to a master equation for the reduced density operator of the DQD in a
many-body basis \cite{StarkEPL2010a}.  Here, the relatively small inter-dot
tunneling in the experiment requires us to work beyond a secular 
approximation, i.e., to take off-diagonal density matrix elements into
account.  For a detailed description of the formalism, see the \supplement.

Quantum dissipation is modeled by a coupling to bosonic environmental 
degrees of freedom with the Hamiltonian $H_\text{el-env} = Z \xi$ with the
(dimensionless) DQD dipole operator $Z = n_L-n_R$ and $n_\ell = 0,1,2$ the
occupation of dot $\ell$, as sketched in Fig.~\ref{fig:broadening}c.
The quantum noise $\xi = \sum_\nu \lambda_\nu(a_\nu^\dagger+a_\nu)$
originates from bosonic modes $\nu$ with frequency $\omega_\nu$,
annihilation operator $a_\nu$, and coupling strength $\lambda_\nu$
\cite{LeggettRMP1987a, HaenggiRMP1990a, Weiss1998a, MakhlinRMP2001a}.  We
assume the modes to be initially at thermal equilibrium with temperature
$T$.  The corresponding rates for dissipation and decoherence can be
expressed in terms of the spectral density $J(\omega) = \pi\sum_\nu
|\lambda_\nu|^2 \delta(\omega-\omega_\nu) \equiv \pi\alpha\omega/2$, which
we assume to be Ohmic, i.e., linear in the frequency $\omega$.  A
particular role is played by the dimensionless dissipation strength
$\alpha$ which determines the magnitude of dissipation and decoherence.
One of our main goals is to determine $\alpha$ from experimental data.
Let us remark that the DQD-bath model has been chosen such that for the DQD
occupation with a single electron, the tunnel term in $H_\text{DQD}$ and
the coupling operator $Z$ can be represented by the Pauli matrices
$\sigma_x$ and $\sigma_z$, respectively.  Then our Hamiltonian becomes the
usual Caldeira-Leggett model \cite{LeggettRMP1987a, HaenggiRMP1990a,
Weiss1998a} $H = \frac{\Omega}{2}\sigma_x + \frac{\epsilon}{2}\sigma_z +
\sigma_z\xi$ for the dissipative two-level system with detuning $\epsilon =
\epsilon_L-\epsilon_R$. It undergoes a phase
transition at $\alpha=1$, while for $\alpha\ll 1$, its dynamics is governed
by quantum coherence.

For a strong detuning of the DQD levels, the dipole operator $Z$ is
practically a good quantum number, i.e., it approximately commutes with the
DQD Hamiltonian and, thus, cannot cause significant transitions.
Therefore, as we will see in our numerical results, $H_\text{el-env}$ may
explain the broadening of the peaks, but not the emergence of the
temperature dependent background.  To model also the latter, we introduce a
second heat bath with the Hamiltonian $H_\text{el-env}' = X \sum_\nu
\lambda_\nu(b_\nu^\dagger+b_\nu)$, where the annihilation operator $b_\nu$, the
spectral density, and the dimensionless dissipation strength $\alpha'$ are
defined as for the first bath. The coupling is established via
the tunnel operator $X = \sum_\sigma (c_{L\sigma}^\dagger c_{R\sigma} +
c_{R\sigma}^\dagger c_{L\sigma})$ and, thus, can induce dissipative
transitions of electrons from one QD to the other.  Hence, this bath is
relevant mainly when the DQD eigenstates are localized, i.e., far from the
peaks where the detuning dominates.  Physically, this corresponds to
coupling the intra-dot current to an environment.

\paragraph{Peak broadening}
The most significant observation in the measured current peaks is their
broadening with increasing temperature.  After subtraction of
the background the peaks reveal a Lorentzian
shape, see Fig.~\ref{fig:broadening}a,b.  In panel b, we also witness that
with increasing temperature, the Lorentzian may be distorted by small
resonances in its vicinity.
Our first goal is to determine the tunnel couplings $\Gamma_{L,R}$ and
$\Omega$ as well as the dimensionless dissipation strength $\alpha$.
Without the background, it turns out to be sufficient to consider only bath~1.
The task is facilitated by the approximate solution for the current
\begin{align}
    I\approx \frac{e}{\hbar}\frac{\Omega^2\big(\gamma-2\pi\alpha\epsilon\big)}
    	    {4\epsilon^2+\big(\gamma + 2\Omega^2/\Gamma_R\big)\gamma}
\label{eq.:IREff}
\end{align}
with $\gamma = \Gamma_R+4\pi\alpha k_BT$ and the detuning $\epsilon = \eta
e(V-V_\text{peak})$ with the peak position $V_\text{peak}$ and the leverage
$\eta=0.15$.  For a derivation, see the \supplement.  Then for the
relatively large temperatures in our experiment, the peak as a function of
the detuning $\epsilon$ has the width $\gamma$, while its height reads
$e\Omega^2/\hbar\gamma$.  Hence, from the linear behavior of the width of
peak I as a function of $T$
(Fig.~\ref{fig:broadening}d), we can immediately read off
$\Gamma_R=\SI{30}{\micro eV}$ and $\alpha = 0.011$.  With these parameters
at hand, the peak height shown in Fig.~\ref{fig:broadening}e provides the
inter-dot tunneling $\Omega = \SI{4.2}{\micro eV}$.  Quite remarkably, the
tunneling from the emitter, $\Gamma_L$, is of minor relevance, as long as
$\Gamma_L\gg\Gamma_R$ showing the dominance of the smallest
rate in tunneling. Using these values, the theoretical results for the
shape of the peaks agrees rather well with the experimental data.

For peak II, the determination of the width at
high temperatures is hindered by the emergence of two small, but close
peaks visible in the inset of Fig.~\ref{fig:broadening}b.  Nevertheless, we
find that the above values for $\Gamma_L$ and $\alpha$ predict also the
width of this peak.  However, matching the peak height requires a
significantly smaller inter-dot tunneling $\Omega = \SI{1.8}{\micro eV}$.
Therefore, we conclude that the two peaks investigated stem from
different DQDs.

\begin{figure}[t]
\centerline{\includegraphics[]{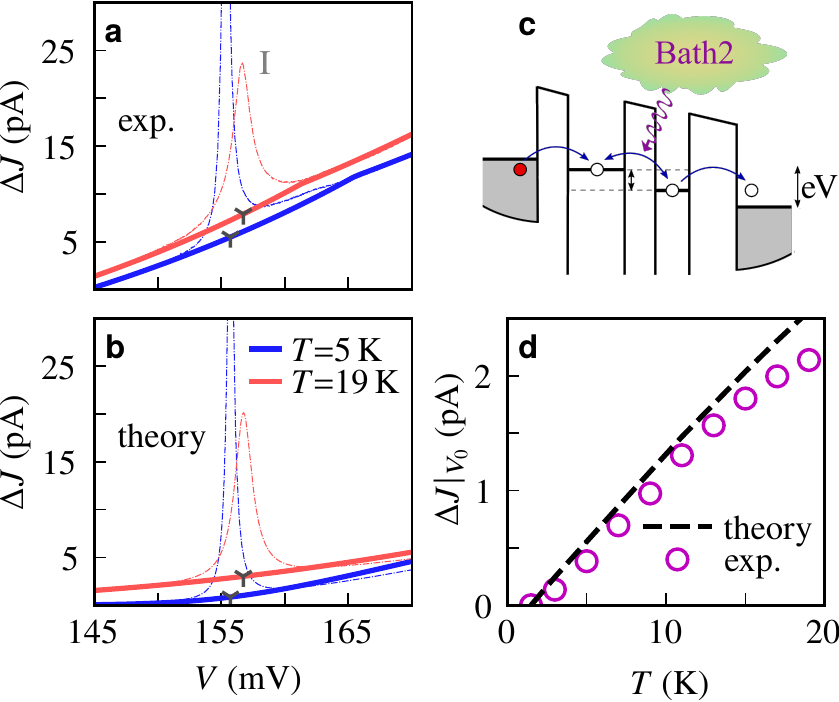}}
\caption{Analysis of the background.
\textbf{a}, Measured tunneling current for the first peak for three
different temperatures (thin black lines) and subtracted background
(coloured lines).
\textbf{b}, Corresponding theoretical results.
\textbf{c}, Sketch of the coupling of the tunnel operator to bath~2 which
causes a smooth background current.
\textbf{d}, shift of the current baseline from its value at \SI{1.5}{K} in
dependence of the temperature, $\Delta J$, at the peak position.  The
theoretical values are computed with $\alpha'=2.4\cdot 10^{-6}$, while all
other parameters are as in Fig.~\ref{fig:broadening}.
\label{fig.:BG}}
\end{figure}

\paragraph{Background}
We have seen that the coupling to a heat bath via the DQD dipole moment can
provide a faithful description of the broadening of the resonance peaks.
However, it does not explain the smooth, temperature-dependent background
witnessed in the experimental data shown in Fig.~\ref{fig.:scheme}a and
detailed in Fig.~\ref{fig.:BG}a.  We attribute this background to a weak
coupling of the inter-dot current to an environment modeled by our second
bath. In the following we estimate the corresponding dissipation strength
$\alpha'$. To compensate the impact of other double dots in our sample, we focus on how the
baseline of the peak raises from its value at the lowest temperature used
in the experiment, $\SI{1.5}{K}$, which provides the data in
Fig.~\ref{fig.:BG}d.  For the theoretical analysis, we now include also the
dissipative inter-dot tunneling sketched in Fig.~\ref{fig.:BG}c.

For the fitting, we again start with an analytic estimate.  In doing so, we
derive with a standard calculation \cite{supplement} the dissipative
transition rate between the states $(1,0)$ and $(0,1)$, which reads
$\kappa(\epsilon) = \pi\alpha'\epsilon/\hbar(1-e^{-\epsilon/kT})$.
Since these dissipative transitions are rather slow, they represent the
bottleneck of the transport and determine the current such that $I \sim
e\gamma(\epsilon)$.  For the dissipation strength $\alpha' =
2.4\cdot10^{-6}$, the theory result for $\Delta J$ agrees with our
experimental data.  Remarkably, already for $\alpha'\sim 10^{-4}\alpha$, the second
bath has a significant influence.  This is in agreement with previous
theoretical findings \cite{BelloPRB2017a} that a heat bath that couples via
the tunnel operator may have a rather
strong impact already for small values of $\alpha'$.

The comparison of the theoretically computed background in
Fig.~\ref{fig.:BG}b exhibit a qualitative agreement with experimental data.
On a quantitative level, however, the shape of the background is reproduced
by the model not as well as the peak itself.  One reason for this is that
the impact of neighboring DQDs cannot be isolated with sufficient
precision.  Another reason is that in contrast to bath~1, the present
dissipative decay probes the spectral density of the bath in a broad
frequency range.  Therefore, the assumption of an Ohmic monotonic spectral density
naturally implies limited agreement \cite{HofmannPRResearch2020a}.
Nevertheless, our model is capable of explaining the physics that leads to
the background of the current peaks.

\begin{figure}[t]
\centerline{\includegraphics[]{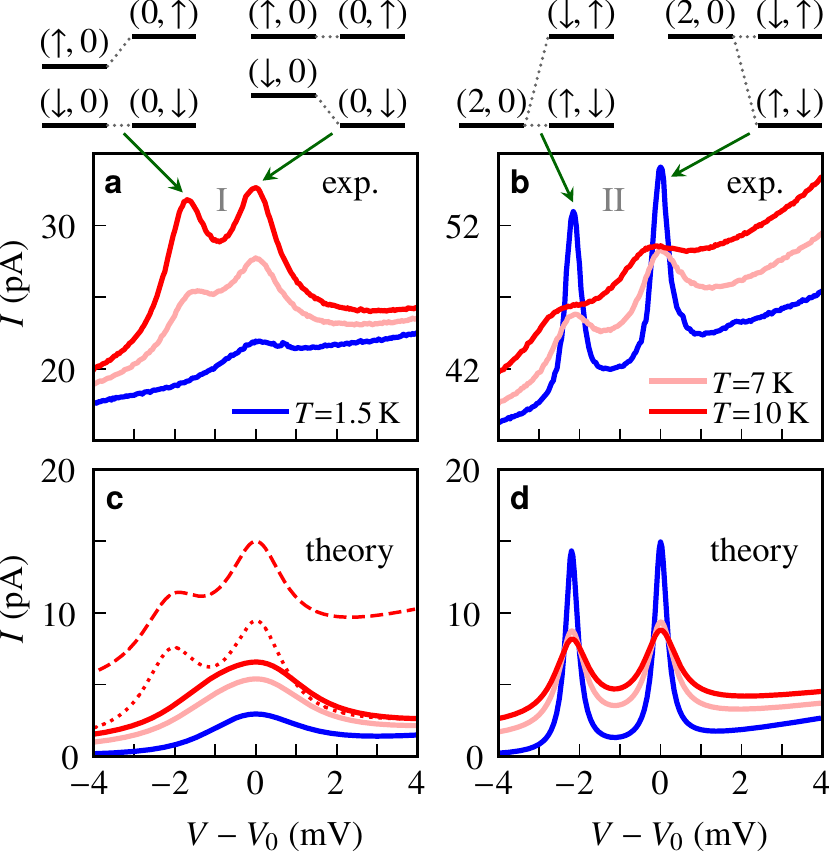}}
\caption{Influence of a magnetic field.
\textbf{a,b}, Temperature dependence of the current peaks at $\SI{155}{mV}$ (a) and
$\SI{188}{mV}$ (b) in a magnetic field with $B=\SI{4}{T}$.
The level schemes on top sketch the resonant processes that lead to the
peaks as is explained in the text.
\textbf{c,d}, 
The theory curves computed with the inhomogeneity $\Delta E_Z =
\SI{0.33}{meV}$, such that for $\eta = 0.15$ the peaks are
separated by \SI{2.2}{mV}.
All other parameters are as in Fig.~\ref{fig.:BG}. The additional lines in
panel c are computed with a significantly larger $\alpha' = 10^{-5}$
(dashed line) and with additional spin flips (dotted, for details see the
\protect\supplement).}
\label{fig.:BField}
\end{figure}

\paragraph{Zeeman splitting}
Both current peaks analyzed above may be fitted with either a one-electron
model or with a model with up to two electrons.  However, as indicated
in Fig.~\ref{fig.:scheme}c,d, both lead to different behavior in the
presence of a magnetic field.  Specifically, peak II
splits up, while peak I does not, at least not at
$\SI{1.5}{K}$. To analyze the impact of a magnetic field in more detail, we
add a Zeeman term to our model and elaborate the resulting temperature
dependence of the peaks.

While in our experiment, the magnetic field is homogeneous, the $g$-factors
of the dots are different due to the different sizes of the two dots and
such the Zeeman splitting becomes inhomogeneous. With the observed splitting of $\Delta V_Z = \SI{2.2}{mV}$ at $B = \SI{4}{T}$, and the leverage factor $\eta=0.15$, we calculate $\Delta g = 1.42$, which seems reasonable in comparison to single dot devices of similar size \cite{Hapke-WurstPhysicaE2002a}.
The details of the individual $g$-factors are not observed in our experiment.  We use in our model as parameters the average Zeeman energy $\bar E_Z$ and their
difference $\Delta E_Z$.  Provided that the Zeeman splitting does not shift
any energy level across the Fermi surface, $\bar E_Z$ has no relevant
influence.  Hence, an appropriate extension of the model can be described
by the Hamiltonian $H_\text{Zeeman} = \frac{1}{2}\Delta E_Z(n_{L\downarrow}
- n_{L\uparrow} - n_{R\uparrow} + n_{R\downarrow})$, where $\Delta E_Z$ is
  determined from the splitting of the peak.

Figure~\ref{fig.:BField} depicts the corresponding measured (panels a and b)
and computed (panels c and d) current peaks for various temperatures.  As
already seen in Fig.~\ref{fig.:scheme}b, at $T=\SI{1.5}{K}$ we witness a
single peak, while Fig.~\ref{fig.:BField}a shows that with increasing
temperature, a second peak emerges.  This observation can be explained with
the level scheme sketched on top of the experimental plots.  For an
inhomogeneous Zeeman splitting, the transport channels cannot be resonant
at the same time.  Hence, an electron may get stuck in the off-resonant
channel and block transport.  However, owing to the coupling to the bath, a
dissipative decay $({\downarrow},0) \to (0,{\downarrow})$ can resolve the
blockade such that a current can flow.  For the spin-up channel, however,
the corresponding process requires the absorption of energy from the bath
and, therefore, can occur only at sufficiently high temperatures.  The
results obtained with our theoretical model qualitatively explain the
emergence of the second peak in the temperature regime of the experiment.
In the numerical data, however, the second peak is poorly resolved, as only
a small shoulder emerges.  For this discrepancy between theory and
experiment two possible reasons come to mind.  First, the magnetic field
may distort the electron wave function such that the dissipation strength
$\alpha'$ becomes larger and, thus, dissipation-assisted tunneling is
enhanced.  The dashed line in Fig.~\ref{fig.:BField}c shows with larger $\alpha'$
indeed a double peak emerges. A further possible explanation is that with increasing Zeeman
splitting, spin flips play a more important role.  Then an electron in the blocked
channel can undergo a spin flip and, thus, end up in the resonant channel.
The dotted line in panel c demonstrates that also this conjecture leads to
a double peak (details of the corresponding model are given the \supplement).

For peak II, this kind of current blockade does not occur, because
the resonant state in the left QD is the spin singlet
$({\uparrow}{\downarrow},0)$ which is unaffected by the magnetic field.
Hence, we observe a double peak also at low temperatures.  With increasing
temperature, each peak broadens much like the single peaks obtained in the
absence of a magnetic field.

\paragraph{Discussion and conclusions}
We have measured resonance peaks in the current through InAs double quantum
dots and showed how their width and background increase with temperature in
the range of 1.5--\SI{21}{K}.  This behavior can be fully explained by
introducing two (bosonic) baths of the Caldeira-Leggett type.  The peak
broadening can be modeled with rather good precision with a bath that  
couples via the dipole moment of the DQD.  Far from the peak, however, this
bath has little influence.  Therefore, we introduced a second bath that
couples to the DQD current which in a tight-binding description is given by
the tunnel operator.  We determined the corresponding dimensionless
dissipation strength of each bath.  Physically, the baths describe the
influence of substrate phonons and also of the impedance of the
electromagnetic environment \cite{DevoretPRL1990a, Ingold1992a}.

A magnetic field makes the situation more complex.  In particular, we found
that then the behavior of the peaks depends on the triple
point at which the DQD is operated.  In turn, this allows one to determine
whether two-electron states play a role.

The interest in the dissipative model parameters extends beyond the
description of current peaks.  For example, with the dot-lead couplings
suppressed, the one-electron states form a charge qubit whose dephasing
time as a function of the dissipation strength is determined by the
corresponding rates in the quantum master equation (see the \supplement)
and in the absence of the second bath is known analytically
\cite{WeissPRL1989a, MakhlinRMP2001a}.  In the temperature regime of the
experiment the present values of $\alpha$ and $\alpha'$ correspond to a
dephasing time of the order $T_2\sim\SI{100}{ps}$, where the precise value
depends also on the detuning.  Assuming that the same model parameters are
valid at lower temperatures, one can predict coherence times up to
\SI{10}{ns}. Therefore, our analysis allows us to predict parameters
being essential for all approaches involving manipulation of quantum
states.

\paragraph{Acknowledgements}
O.D., J.C.B., and R.J.H.\ acknowledge funding by the Deutsche
Forschungsgemeinschaft (DFG, German Research Foundation) under Germany's
Excellence Strategy -EXC 2123 QuantumFrontiers-390837967 and the State of
Lower Saxony of Germany via the Hannover School for Nanotechnology.  S.K.
acknowledges financial support by the Spanish Ministry of Science and
Innovation through Grant.\ No.\ PID2020-117787GB-I00 and the CSIC Research
Platform on Quantum Technologies PTI-001. The authors thank Jan K\"uhne and
Felix Opiela for contributions at the beginning of the project.

\paragraph{Author contributions}
O.D. and J.C.B. performed the measurements and the analysis of the
experimental data.  R.H. and S.K. developed the theoretical model and
computed the numerical data.  All authors participated in the discussions
of the results and contributed to the writing and editing the manuscript.
The research was supervised by S.K. and R.J.H.

\paragraph{Competing interests}
The authors declare no competing interests.

\paragraph{Methods}

\subparagraph{Experiment}
The experimental device consists of self-assembled InAs double quantum
dots, stacked between GaAs leads with annealed metal contacts. Three AlAs layers of
nominally identical thickness $\SI{5}{nm}$ are used to separate the two
InAs quantum dot layers from the (doped) GaAs leads and from each other.
Temperature and magnetic field control
were achieved by placing the device into a He4 cryostat with variable
temperature insert. For all measurements, the bias voltage was applied to
the contact closer to the layer of smaller QDs (right). The contact closer
to the larger QDs (left) was connected to a current preamplifier
(\SI{1e-10}{\ampere\per\volt}) to ensure high sensitivity. Each datapoint
was obtained by integrating the voltage output of the current preamplifier
over one power line cycle ($\SI{20}{\milli\second}$).

\subparagraph{Bloch-Redfield master equation}
The full model Hamiltonian for the DQD and its environment is of the
structure $H = H_\text{DQD} + H_\text{env} + V$, where $H_\text{env}$ and
the DQD-environment coupling $V$ contain a summation over all heat baths
and leads.  The dynamics of the total density operator is governed by the
Liouville-von Neumann equation
$\dot\rho_\text{tot} = -i [H,\rho_\text{tot}]$ which is practically
intractable owing to its macroscopic number of degrees of freedom.
Therefore, we derive a master equation for the density operator of the DQD
by integrating out the environment within second-order perturbation theory,
which in units with $\hbar=1$ reads
\cite{RedfieldIBMJRD1957a, Breuer2003a, StarkEPL2010a}
\begin{equation}
\dot\rho = -i[H_\text{DQD},\rho]
- \int_0^\infty d\tau \tr_\text{env}
  [V,[V(-\tau),\rho\otimes\rho_\text{env}]] ,
\label{BR}
\end{equation}
where $V(t) = e^{i(H_\text{DQD} + H_\text{env})t} V e^{-i(H_\text{DQD} +
H_\text{env})t}$ is the coupling operator in the interaction picture and
$\tr_\text{env}$ the trace over all bath and lead variables.  Generally,
Eq.~\eqref{BR} possesses a unique stationary solution which allows us to
compute the current.

The numerical solution of the master equation is conveniently performed in
the energy eigenbasis of the DQD with the many-body eigenstates $|k\rangle$ and the
corresponding energies $E_k$ and occupation numbers $N_k$.  The main
advantage of this representation is that it brings $V(t)$ to a simple form
such that the time integration in Eq.~\eqref{BR} can be evaluated
analytically.  The direct transitions between the populations are
determined by golden-rule rates.  For example, tunneling of an electron
from a lead to the DQD occurs with a rate $\Gamma^\text{(in)}_{kl}\propto
f(E_f-E_i-\mu)$ with $f$ being the Fermi function and $E_i$ and $E_f$ the
energies of the initial and final DQD state, whose difference must be
compensated by the energy of the incoming lead electron.  Hence, these
terms can occur only when $E_f \lesssim E_i+\mu$.  In turn,
$\Gamma^\text{(out)}_{kl}\propto f(E_f-E_i+\mu)$, which differs from
$\Gamma^\text{(in)}$ only by the sign of the chemical potential.  For the
dissipative transitions, one finds absorption and emission rates which are
linked by Boltzmann factors.

Importantly, owing to the relatively small inter-dot tunneling in our
system, off-diagonal density matrix elements are rather relevant.  Indeed, one
observes that $\rho$ eventually becomes block diagonal in the occupation
number $N_k$, where the blocks correspond to subspaces with equal DQD
occupation \cite{DarauPRB2009a}.  Within these blocks, however, off-diagonal density matrix
elements may have an appreciable size, which implies that a secular
approximation is suitable only for pairs of states with different
occupation number.

\cleardoublepage
\setcounter{section}{0}
\renewcommand{\thefigure}{S\arabic{figure}}
\setcounter{figure}{0}

\section*{Supplemental Material}

\section{Additional data}

\begin{figure}[t]
\includegraphics{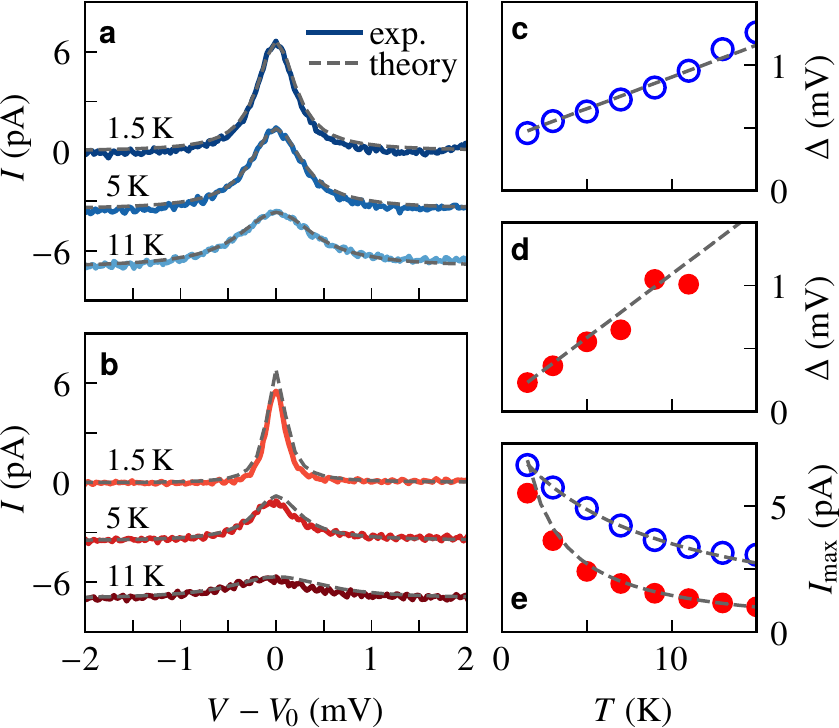}
\caption{(a,b) Current peaks measured at a further sample for various
temperatures.  (c,d,e) Corresponding peak widths and heights plotted with
the respective colour.
The dashed lines are theory data for the parameters summarized in
Table~\ref{table:params}.
\label{fig:additional}
}
\end{figure}

In Fig.~\ref{fig:additional}, we show experimental data together with
theory results for two peaks measured with a different sample of the
similar layer structure. The only difference in the layer structure is a
slightly increased thickness ($\SI{7}{nm}$) of the AlAs barrier layer
separating the two InAs dot layers. All other measurement conditions were
kept identical to the data discussed in the main text. In the presence of a
magnetic field at \SI{1.5}{K} (not shown), only the peak in panel b
splits and, thus, the corresponding triple point must be of type II.  By
contrast, the behavior of the peak in Fig.~\ref{fig:additional}a indicates
a triple point of type I.  The widths and heights of the peaks can be
appreciated in Figs.~\ref{fig:additional}c,d,e.  For temperatures up to
\SI{10}{K}, they are well reproduced by our theory.  For larger
temperatures, the peaks become very small, such that height and width
cannot be determined with sufficient precision.

\section{Data analysis}

The data analysis relies on the assumption that the background left and
right to a resonance peak of the experimental current-voltage
characteristics is growing exponentially. Their slopes are, however,
distinct. It consists of (i) the subtraction of the left and right
background from the raw data and (ii) the simultaneous fitting of the
extracted data for different temperatures with the formula given
in Eq.~\eqref{eq.:IREff} of the main text.

In step (i), we first estimate the position and line width of the resonance 
in order to separate the voltage regimes relevant for the left and right
background. Then, we determine their exponential regressions. Finally, we
identify their intersection at which we stitch both backgrounds together.
The combined background is then subtracted from the measured
current-voltage characteristics. This scheme is applied to all discussed
datasets and for each temperature. For a fixed dataset, we fit the
extracted current-voltage characteristics simultaneously for different
temperatures using averaged parameters from individual fits as initial
guess. The resulting fitting parameters serve as starting point for the
theoretical description incorporating the spin as discussed in the following section.

\begin{table}[t]
\centering
\caption{Parameters for the theory data shown in
Fig.~\ref{fig:additional}.  The emitter-DQD rate, $\Gamma_L =
\SI{0.5}{meV}$, is the same as in the main text.
\label{table:params}}
\begin{ruledtabular}
\begin{tabularx}{\columnwidth}{lccccc}
Figure & Type of TP & $\Omega$ & $\Gamma_R$ & $\alpha$ & $\alpha'$ \\
\colrule
\ref{fig:additional}a & I    & \SI{1.4}{\micro eV} & \SI{60}{\micro eV} & 0.007 & $2.2\cdot 10^{-6}$ \\
\ref{fig:additional}b & II   & \SI{0.7}{\micro eV} & \SI{12}{\micro eV} & 0.014 & $2.2\cdot 10^{-6}$ \\
\end{tabularx}
\end{ruledtabular}
\end{table}

\section{Three-level approximation}

In refs.~\cite{KiesslichPRL2007a, SanchezPRB2008a}, a similar model has been
used, but without taking the spin degree of freedom into account.  Then in
the limit of strong interaction, double occupancy is energetically
forbidden and the master equation possesses an analytic solution.  Our goal
is now to approximate our model in the vicinities of triple points by a
three-level model and to show that the resulting Liouvillian can be
approximated by the one for spinless electrons.  In doing so, we obtain an
approximative analytical solution which facilitates parameter fitting.

\subsection{Analytical solution for spinless electrons}

When disregarding spin and double occupation, the DQD Hamiltonian can be
decomposed into the basis states $|0\rangle$, $|L\rangle$, $|R\rangle$,
which refer to the empty DQD and the occupation with one electron on the
left and the right dot, respectively.  Since in the stationary limit,
coherences between the state $|0\rangle$ and the one-electron states
vanish, the DQD density operator can be written in the basis of the
populations $p_0$, $p_L$, and $p_R$, together with the coherences
$\rho_{LR}$ and $\rho_{RL}$.  The corresponding Liouvillian reads
\cite{SanchezPRB2008a}
\begin{equation}
\mathcal{L} =
\begin{pmatrix}
-\tilde\Gamma_\text{in} & 0 & \tilde\Gamma_\text{out} & 0 & 0 \\
 \tilde\Gamma_\text{in} & 0 & 0     & -\frac{i}{2}\tilde\Omega & \frac{i}{2}\tilde\Omega\\
 0 & 0 & - \tilde\Gamma_\text{out}  &  \frac{i}{2}\tilde\Omega & -\frac{i}{2}\tilde\Omega\\
 0 &  -\frac{i}{2}\tilde\Omega+A_+ &\frac{i}{2}\tilde\Omega-A_- & i\tilde\epsilon-B & 0\\
 0 &   \frac{i}{2}\tilde\Omega+A_+ & -\frac{i}{2}\tilde\Omega-A_- & 0 & -i\tilde\epsilon-B\\
\end{pmatrix} ,
\label{Lspinless}
\end{equation}
where
\begin{align}
A_{\pm} ={}& \frac{\pi}{2}\tilde\alpha\tilde\Omega
\pm \frac{\pi}{2}\tilde\alpha\tilde\epsilon\tilde\Omega \Big(\frac{2k_BT }{E^2}
- \frac{1}{E}\coth\frac{E}{2k_BT }\Big) ,
\\
B ={}& \pi\tilde\alpha\Big(\frac{2\tilde\epsilon^2k_BT}{E^2}
	+ \frac{\tilde\Omega^2}{E} \coth\frac{E}{2k_BT}\Big)
	+ \frac{\tilde\Gamma_\text{out}}{2} ,
\end{align}
and $E = (\tilde\epsilon^2+\tilde\Omega^2)^{1/2}$ is the splitting of the
one-electron states.

The corresponding stationary density operator and, thus, the corresponding
current can be calculated analytically,
\begin{align}
\label{Ispinless0}
I ={}&\frac{e}{\hbar}\frac{\tilde \Omega^2\big(2B-4 A_+\tilde\epsilon/\tilde\Omega\big)}{
	4(\tilde\epsilon^2+B^2) +\tilde\Omega^2\Big(
		\frac{4B+4(A_- -A_+)\tilde\epsilon/\tilde\Omega}{\tilde\Gamma_\text{out}} 
		+ \frac{2B-4 A_+\tilde\epsilon/\tilde\Omega}{\tilde\Gamma_\text{in}} 
	\Big)
}\\
\approx{}&\frac{e}{\hbar}\frac{\tilde \Omega^2\big(\gamma-2\pi\tilde\alpha\tilde\epsilon)}{
	4\tilde\epsilon^2+\gamma^2 +\tilde\Omega^2\Big(
		\frac{2\gamma}{\tilde\Gamma_\text{out}} 
		+ \frac{\gamma-2\pi\tilde\alpha\tilde\epsilon}{\tilde\Gamma_\text{in}} 
	\Big)
}
\label{Ispinless1}
\\
\approx{}&\frac{e}{\hbar}
\frac{\tilde\Omega^2\big(\gamma-2\pi\tilde\alpha\tilde\epsilon\big)}
     {4\tilde\epsilon^2+\big(\gamma + 2\tilde\Omega^2/\tilde\Gamma_\text{out}\big)\gamma},
\label{Ispinless}
\end{align}
where $\gamma = \tilde\Gamma_\text{out} + 4\pi\tilde\alpha k_BT$.  The
approximation in Eq.~\eqref{Ispinless1} holds for sufficiently high
temperature, such that in $A_\pm$ and $B$, the hyperbolic cotangent becomes
$\coth(E/2k_BT) \approx 2k_BT/E$. The final expression \eqref{Ispinless}
holds if, in addition, $\tilde\Gamma_\text{in} \ll\gamma$, which in our
setup is ensured by the fact that the emitter barrier is noticeably thinner
than the collector's one.

\subsection{Effective parameters}

\begin{table}[t]
\centering
\caption{Mapping of the three-level approximation to a model with spinless
electrons.  The corresponding parameters for the detuning $\tilde\epsilon$,
the tunnel coupling $\tilde\Omega$, the dot-lead rates
$\tilde\Gamma_\text{in,out}$, and the dissipation strengths $\tilde\alpha$
and $\tilde\alpha'$ depend on the triple point at which the DQD is
operated.
\label{tab:mapping}}
\begin{ruledtabular}
\begin{tabularx}{\columnwidth}{lcccccc}
Three-level model & $\tilde\epsilon$ & $\tilde\Omega$ & $\tilde\Gamma_\text{in}$
	& $\tilde\Gamma_\text{out}$
& $\tilde\alpha$ & $\tilde\alpha'$ \\
\colrule
TP I	& $\epsilon$	& $\Omega$		& $2\Gamma_L$	& $\Gamma_R$
	& $\alpha$	& $\alpha'$ \\
TP II	& $\epsilon$	& $\sqrt{2}\Omega$	& $\Gamma_L$	& $\Gamma_R$
	& $\alpha$	& $2\alpha'$ \\
TP III	& $\epsilon$	& $\Omega$		& $\Gamma_R$	& $2\Gamma_L$
	& $\alpha$	& $\alpha'$ \\
\end{tabularx}
\end{ruledtabular}
\end{table}

The mapping of our system to the Liouvillian \eqref{Lspinless} is achieved
by two steps. First, one selects the three electron configurations, which may imply
a summation over occupation probabilities for different spin
configurations.  Second, one relates the parameters of the full model to
those of the spinless model, such that the current can be inferred from
Eq.~\eqref{Ispinless}.  The results are summarized in
Table~\ref{tab:mapping}.  For all cases, we have verified numerically that
in the relevant regime, the mapping represents a rather good approximation
to the full description.

All triple points have in common that the coherent inter-dot tunneling
connects two states which differ in the position of one electron.
Therefore, the detuning $\epsilon$ has the same effect for all cases.
Moreover, the transition matrix element of the dipole operator at all
triple points agrees with the one of the spinless model.  Hence, the
dissipation strength of bath 1, which couples via the dipole operator, is
the same in all cases, $\tilde\alpha=\alpha$.

\subsubsection{Triple point I}

Triple point I connects the empty DQD state and the occupation with one
electron on the left or the right dot.  Thus, the situation comes close to
the spinless case.  The main difference is that an electron that arrives
from the emitter may have any spin, such that the summation over the spin
degree of freedom provides a factor 2 for the corresponding rate,
$\tilde\Gamma_\text{in} = 2\Gamma_L$ (in a more formal treatment, one
defines a probability $p = p_\uparrow+p_\downarrow$ \cite{HusseinPRL2020a}).
For all subsequent transitions, including the dissipative ones, the spin is
conserved such that all other parameters equal those of the spinless case.

\subsubsection{Triple point II}

At triple point II, the state with lowest occupation has one electron on
the dot connected to the emitter.  Owing to the Pauli principle, a
second electron that tunnels to the left dot must have opposite
spin.  Therefore, no factor 2 emerges such that $\tilde\Gamma_\text{in} =
\Gamma_L$.  For the tunneling from the DQD to the collector, the situation is
the same as at triple point I.

The more interesting issue is that the $(1,1)$ configuration may come as
singlet or triplet, while in our model with one orbital per dot, the
$(2,0)$ state is necessarily a singlet (the triplet state is higher in
energy and, thus, is not accessible).  Therefore, coherent tunneling occurs
between the singlets $|{\uparrow\downarrow},0\rangle$ and
$(|{\uparrow},{\downarrow}\rangle+
|{\downarrow},{\uparrow}\rangle)/\sqrt{2}$.  The corresponding matrix
element of the inter-dot tunneling is $\sqrt{2}\Omega$, which can be
identified with the tunneling in the spinless model.  Dissipation in
principle may cause leakage to a $(1,1)$ singlet, which would not be
covered by our three-level model.  In practice, however, these transitions
do not play a role, because they are slower than the tunneling to the
collector.

The factor $\sqrt{2}$ also occurs for the coupling to bath 2, which is
established via the tunnel operator.  Since the dissipative rates are
proportional to the square of the transition matrix elements, one finds the
correspondence $\tilde\alpha' = 2\alpha'$.

\subsubsection{Triple point III}

A striking feature of triple point III is that, in contrast to TP~I and TP~II, it
is traversed counter-clockwise.
Then the coherent inter-dot tunneling is initialized by the tunneling of an
electron from the singly occupied right dot to the collector, such that we have
the correspondence $\tilde\Gamma_\text{in} = \Gamma_R$.  After the inter-dot
tunneling, the system is reset by an electron tunneling from the left lead
to the empty left dot.  Hence, $\tilde\Gamma_\text{out} = 2\Gamma_L$.
A shortcut to this reasoning can be established upon noticing that the
situation can be interpreted as hole transport from the collector to the
emitter.  Then we are back to the situation at TP~I, but with
$\tilde\Gamma_\text{in}$ and $\tilde\Gamma_\text{out}$ interchanged.

\subsubsection{Triple point IV}

Let us finally remark that triple point IV is fundamentally different from
the other three triple point, because emitter-DQD tunneling $(0,1) \to
(1,1)$ may directly create a spin triplet which is not tunnel coupled to
the $(0,2)$ singlet.  This leads to Pauli spin blockade \cite{WeinmannPRL1995a,
OnoScience2002a} which can be lifted by dissipative processes such as spin dephasing
or spin flips that may cause some residual currents.  Such dissipative spin
dephasing or spin flips, however, are fundamentally different from coherent
inter-dot tunneling which dominates in the spinless model (and at the other
triple points).  Consequently, the dynamics at TP~IV cannot be captured by
a spinless model.

\section{Analytical expression for the background}

To understand the action of bath 2, we focus on the regime aside the peaks
in which the DQD eigenstates are well approximated localized states.  There
the bath coupling $H' = X \sum_\nu \lambda_\nu(b_\nu^\dagger+b_\nu)$ with $X
= \sum_\sigma (c_{L\sigma}^\dagger c_{R\sigma} + c_{R\sigma}^\dagger
c_{L\sigma})$ induces dissipative transitions between states with different
charge configuration, which enables transport.  At triple point~I, these
configurations are $(1,0)$ and $(0,1)$.  The transition rate can be
calculated directly from Fermi's golden rule or by evaluating the
corresponding terms of the master equation \eqref{BR} and reads
\begin{equation}
\kappa_\text{I}(\epsilon)
= \frac{\pi\alpha' \epsilon/\hbar}{1-e^{-\epsilon/kT}} .
\label{gamma}
\end{equation}
The expression is the product of the spectral density at the transition
energy $\epsilon = E_{(1,0)} - E_{(0,1)}$ and the corresponding bosonic
thermal occupation number.  The resulting absorption rate follows formally
by inverting the sign of the energy difference.  Since $\kappa_\text{I}$
is much smaller than the dot-lead rates $\Gamma_{L,R}$, the transitions
induced by bath 2 represent the bottleneck of the transport process. Hence,
the resulting current is approximately $I \sim e\kappa_\text{I}(\epsilon)$.

As discussed above, the tunnel matrix element between the singlets with
charge configurations $(2,0)$ and $(1,1)$ augments the effective dissipative
tunnel rate by a factor 2. Therefore, at triple point II, $I\sim
e\kappa_\text{II}(\epsilon)$ with $\kappa_\text{II}(\epsilon) =
2\kappa_\text{I}(\epsilon)$.

\section{Additional spin flip noise}

A Zeeman splitting enables additional dissipative processes such as spin
flips.  They may contribute to the resolution of the blockade mechanism at
triple point I discussed in the main text.  While a detailed investigation
of this phenomenon is beyond the scope of the present work, we nevertheless
provide some data to demonstrate that the role of spin flips may explain
the discrepancy between theory and experiment at triple point I in the
presence of a magnetic field.

We use a model similar to the one of ref.~\cite{DiVincenzoPRB2005a} with a
bosonic bath that couples to one particular spin component.  We choose the
operator $\sigma_x$ of each electron, which flips its spin.  In second
quantization, the corresponding system bath operator reads
\begin{equation}
V = \sum_{i=L,R} ( c^\dagger_{i\uparrow} c_{i\downarrow} 
+ c^\dagger_{i\downarrow} c_{i\uparrow} )\xi_i ,
\end{equation}
where the $\xi_i$ is a Ohmic bath operator acting on the spin of dot $i$.
They are defined as in the main text and have equal dimensionless coupling
strength $\alpha_\text{spin}$.

\section{Dephasing time of a charge qubit}

\begin{figure}[b]
\centerline{\includegraphics{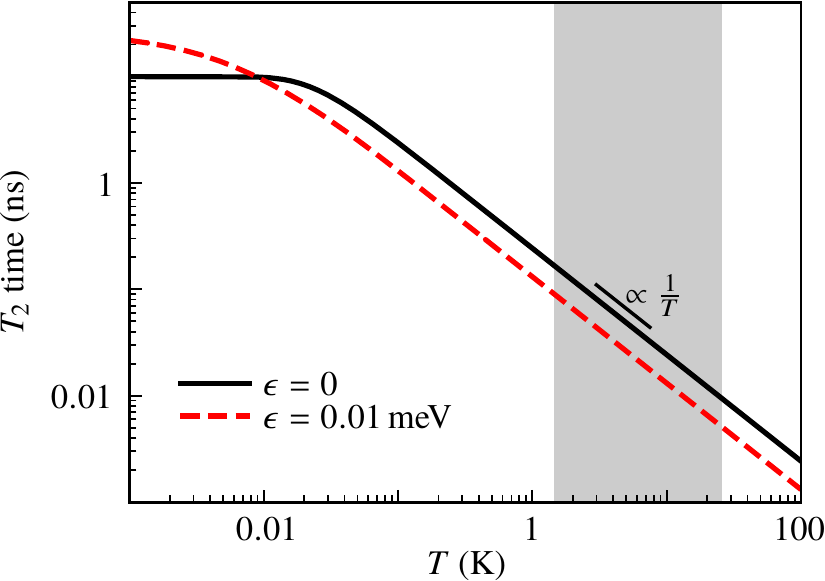}}
\caption{Dephasing time $T_2$ of our model with the dimensionless
dissipation strength determined in the main text.  The values depict the
inverse of the decay rate of the off-diagonal matrix element of the
eigenstates that form the qubit at triple point I.  All parameters are
those determined by fitting to the current peaks in the main text.
The grey area marks the temperature range of the experiment.}
\label{fig:T2}
\end{figure}

The dissipative dynamics of a qubit possesses two characteristic times.
While the population of the excited state decays during a relaxation time
$T_1$, coherent oscillations and interferences fade away exponentially on
a time scale $T_2$.  For the Caldeira-Leggett model with only bath 1, i.e.\
for $\alpha'=0$, both $T_1$ and $T_2$ can be obtained analytically and read
\cite{Weiss1998a, MakhlinRMP2001a}
\begin{align}
\label{T1}
T_1^{-1} ={}& \frac{\pi\alpha}{\hbar} \frac{\Omega^2}{E}
              \coth\Big(\frac{E}{2kT}\Big) ,
\\
\label{T2}
T_2^{-1} ={}& \frac{1}{2}T_1^{-1} + \frac{\pi\alpha}{\hbar}
              \frac{2kT\epsilon^2}{E^2} ,
\end{align}
with the energy splitting $E=\sqrt{\Omega^2+\epsilon^2}$.
In the high-temperature limit, $kT\gg E$, both times are
proportional to the inverse temperature $1/T$, while they saturate at low
temperature.

In principle, Eqs.~\eqref{T1} and \eqref{T2} overestimate the coherence
time of our charge qubit based on one- and two-electron states in a DQD as
it uses a two-level approximation and neglects bath 2 and virtual
transitions to the leads.  To include the second bath, we consider the
dissipative terms of our Bloch-Redfield master in the energy basis.  This
allows us to directly read off the decay rates of the populations and the
coherences.  In contrast to our transport calculations, we here disregard
the dot-lead couplings and consider a qubit in a closed DQD configuration.

In Fig.~\ref{fig:T2} we compare the coherence time $T_2$ at triple point~I
as a function of temperature obtained from our master equation for various
detunings.  For the rather low value of $\alpha'$ in our setup, the result
agrees almost perfectly with Eqs.~\eqref{T1} and \eqref{T2}.  For
temperatures larger than $\sim\SI{20}{mK}$, we witness the $1/T$ behavior
of the high-temperature limit.

\end{document}